# Size dependence of the surface tension of a free surface of an isotropic fluid

**Sergii Burian[1], Mykola Isaiev[1,2], Konstantinos Termentzidis[2], Vladimir Sysoev[1] and Leonid Bulavin[1]**

[1]*Faculty of Physics, Taras Shevchenko National University of Kyiv, 64/13, Volodymyrska Street, Kyiv, 01601, Ukraine*

[2]*LEMTA - Université de Lorraine - CNRS UMR 7563, Faculté des Sciences et Technologies, Boîte Postale 70239, 54506 Vandoeuvre les Nancy cedex, France*

## Abstract

We report on the size dependence of the surface tension of a free surface of an isotropic fluid. The size dependence of the surface tension is evaluated based on the Gibbs-Tolman-Koenig-Buff equation for positive and negative values of curvatures and the Tolman lengths. For all combinations of positive and negative signs of curvature and the Tolman length, we succeed to have a continuous function, avoiding the existing discontinuity at zero curvature (flat interfaces). As an example, a water droplet in the thermodynamical equilibrium with the vapor is analyzed in detail. The size dependence of the surface tension and the Tolman length are evaluated with the use of experimental data of the International Association for the Properties of Water and Steam. The evaluated Tolman length of our approach is in good agreement with molecular dynamics and experimental data.

## 1. Introduction

The size-dependence phenomena play a predominant role for systems with an important surface-to-volume fraction. Therefore, the prediction and tuning of the behavior of interfaces between different phases are a big challenge for various applications such as catalysis in chemistry, biomedical applications, nanostructure formation, etc. [1,2]. Several applications require an in-depth elucidation of the nucleation and the evolution of droplets and bubbles at the nanometer scale [3,4]. For the correct description one needs to know the surface tension and its size and temperature dependence. The size- dependence of the surface tension $\sigma(R)$ can be defined according to the Tolman approximation ($\delta << R$):

$$\sigma(R) = \frac{\sigma^{\infty}}{1+(n-1)\frac{\delta}{R}}, \qquad (1)$$

where $\sigma^{\infty}$ is the surface tension of a flat surface, $n$ is a parameter equal to 3 and 2 for spherical and cylindrical surfaces respectively, $R$ is the radius of the principal curvature of surface of tension, end $\delta$ is the Tolman length's characteristic [5]. The $\delta$ is the difference between the surface of tension and the equimolar surface and as it can be seen in Eq. (1) it defines the change rate of the surface tension with the size. It is clear that the Tolman length is an important parameter for the prediction of nanoscale objects' behavior, especially for processes of nucleation [6–8] and the droplet formation [1,9–12].

There are several approaches to evaluate the Tolman length; among them one can note thermodynamical [13], statistical [14], density functional methods [15–17], and molecular dynamics [18–20], etc. However, with these techniques the Tolman length can be extracted only from the geometry of the interfacial region. For the simplest case of a free isotropic fluid there are only two possibilities of equilibrium shapes, namely, the spherical and cylindrical interfaces [21,22].

For the macroscopic description, the finite interface region can be regarded as a surface – a surface of discontinuity. The physical quantities crossing this surface of discontinuity can be expected as fast-changing things. Nevertheless, for the nanoscopic description the consideration of the finite thickness (surface of discontinuity) of the liquid-vapor interface is crucial. Respectively, the physical quantities in this region change continuously from the bulk value of the first half-space to another [23–25]. For example, in the case of the liquid-vapor dividing surfaces the density changes from the liquid bulk density ($\rho^{\ell}$) to the concentration of the saturated vapor ($\rho^{\upsilon}$). Under the term "saturated vapor", we consider the equilibrium state of liquid and vapor phases. For the case of a curved surface in the literature "the saturation vapor pressure over a curved surface" [26,27] is also often used. The sketch view of the real system is presented in Fig. 1. In the upper-left and lower-left parts of Fig. 1, the schematic distribution of particles in the system and the density distribution in the Helmholtz representation are shown. The red and the blue lines separate liquid and vapor phases respectively. The surface of discontinuity is the space between these lines.

The Gibbs approach is based on the introduction of a reference system with the arbitrary chosen sharp interface, the so-called dividing surface even for nanoscopic objects. [4] The image on the right side of Fig. 1 illustrates the two cases of the dividing surface setting, when a dividing surface serves as an equimolar surface (upper side) and a surface of tension (lower side). The insets show the schematic diagram of the density's spatial distribution for the considered systems. The physical quantities in the subsystems divided by this interface are set the same as the corresponding bulk values in the real system. The introduction of the reference system allows one to avoid issues related with the spatial distribution of the physical quantities inside the interface region. However, the number of particles is the same for both real and reference systems only in the case when the dividing surface between two phases is the equimolar surface (see Fig. 1). In such consideration, the Tolman length is a distance from the equimolar surface to the surface of tension [5].

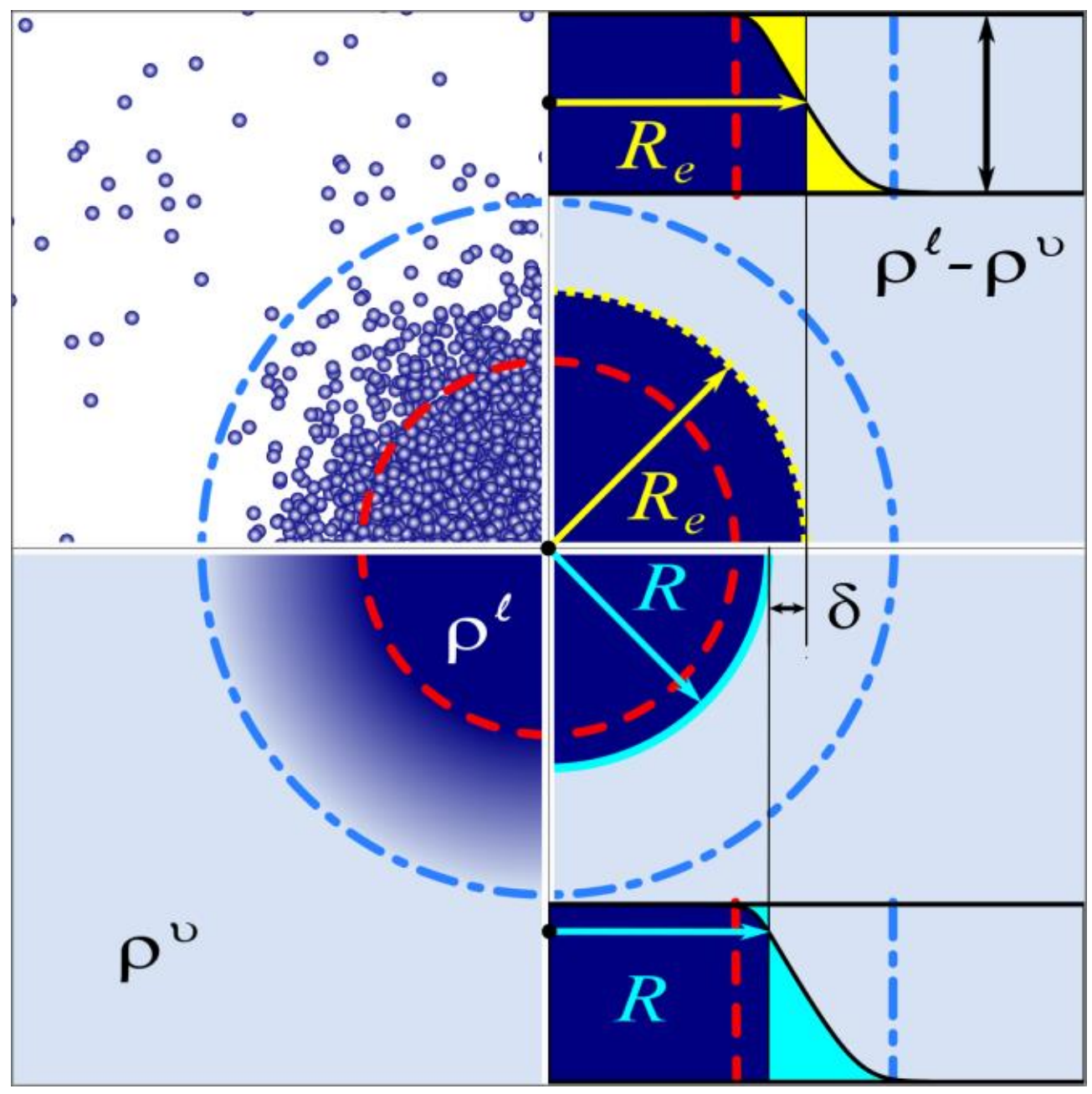

**Fig. 1.** Sketch representation of the real (left) and reference system (right). The dashed and the dashed-dotted lines separate the bulk liquid and the bulk vapor phases, respectively. The surface of discontinuity is the space between these lines. The surface of tension and the equimolar surface is noted by solid and dotted lines, respectively.

Previously, Rakhviashvili and Kishtikova [28–32] proposed an analytical approach for the size dependence of the surface tension with the use of the Tolman length. This approach assumes that the Tolman length is equal to the thickness of the region that corresponds to the distance between the two bulk phases (liquid and vapor). However, this assumption can give only an upper estimate of the Tolman length [33]. Their proposed approach can be applied only when $\delta/R$ is positive or negative and not at $\delta/R = 0$, where there is a discontinuity. This discontinuity leads to limitations of the methodology's applicability [1,18,20,34,35].

The goal of the paper is to study the temperature dependence of the Tolman length of the free surface of tension. We use the method of Gibbs dividing surfaces as an efficient tool for the description of the size dependence of the surface tension. Thus, we propose an approach for evaluation of the Tolman length for different geometries of the free surface of tension. Our approach allows distinguishing the terms "thickness of the interfacial region" and "the Tolman length". For this purpose, we consider an isotropic two-phases system – a liquid in thermodynamic equilibrium with saturated vapor using experimental data of the International Association on the Properties of Water and Steam (IAPWS).

## 2. Methodology

With the use of Gibbs adsorption and Gibbs fundamental equations [4,23,24,36], one can obtain a joint integral equation [37] defining the size dependence of the surface tension $\sigma_{s,c}(R)$ on the radius of curvature $R$:

$$\ln\frac{\sigma_{s,c}(R)}{\sigma^{\infty}}=\int_{\infty}^{R}\frac{\left(\left(1+\delta_{s,c}/r\right)^{n}-1\right)(n-1)/n}{1+\left(\left(1+\delta_{s,c}/r\right)^{n}-1\right)(n-1)/n}\frac{dr}{r}, \tag{2}$$

where $\sigma^{\infty}=\lim_{R\to\infty}\sigma(R)$ is the surface tension of a flat surface, $n=3$ for spherical (s) and $n=2$ for cylindrical (c) surfaces. In the particular case $n=3$, Eq. (2) is the same as the Gibbs [24]–Tolman [5]–Koenig [38]–Buff [39] (GTKB) equation.

For a spherical surface of tension ($n=3$) the analytical solution can be presented as follows:

$$\frac{\sigma_{s}(R)}{\sigma^{\infty}}=\left|\frac{z}{z-\delta/R}\right|^{\frac{3+3z+z^{2}}{3(1+z)^{2}}}\left(\frac{x^{2}+y^{2}}{\left(x+\delta/R\right)^{2}+y^{2}}\right)^{a}e^{2b\left(\arctan\left(\frac{x}{y}\right)-\arctan\left(\frac{x+\delta/R}{y}\right)\right)}, \tag{3}$$

where $x=1-1/2\sqrt[3]{2}$, $y=\sqrt{3}/2\sqrt[3]{2}$, $z=-1-1/\sqrt[3]{2}$, $a=\left(2-\sqrt[3]{2}\right)\left(1+\sqrt[3]{2}\right)/6$, $b=\left(1+\sqrt[3]{2}\right)/\sqrt[3]{4}\sqrt{3}$. The solution for the spherical droplet was presented earlier in [9]. However, Eq. (3) has a more compact form.

The dependence of the surface tension as a function of $\delta/R$ for spherical geometry evaluated with Eq. (3) is presented in Fig. 2(a). As one can see in the figure, the dependence is a continuous real function. As it was mentioned previously the solution presented in Refs. [28–30] has a discontinuity in the point $\delta/R=0$ (see dashed and dash-dotted lines in Fig. 2(a)). Thus, their solution can be applied only when $\delta/R>0$ [28,29] or when $\delta/R<0$ [30]. Additionally, it should be noted that Eq. (3) gives the same results (see Fig. 2a) as the solution presented [9].

For a cylindrical surface of tension ($n=2$) the solution of Eq. (2) can be written as follows:

$$\frac{\sigma_{c}(R)}{\sigma^{\infty}}=\sqrt{\frac{2}{1+\left(1+\delta/R\right)^{2}}e^{\frac{\pi}{4}-\arctan(1+\delta/R)}}. \tag{4}$$

The presented above equation is original, and it is applicable for all combinations of signs of the curvature and Tolman length. The dependence of the surface tension on $n=2$ for the cylindrical geometry evaluated with Eq. (4) is presented in Fig. 2(b). In this case the function is also continuous, and the dependence can be applied for the description of both cases, $\delta/R>0$ and $\delta/R<0$. For the correlation, the dependence obtained earlier for the case of a cylindrical droplet [29] is also presented in Fig. 2(b) by the dashed line.

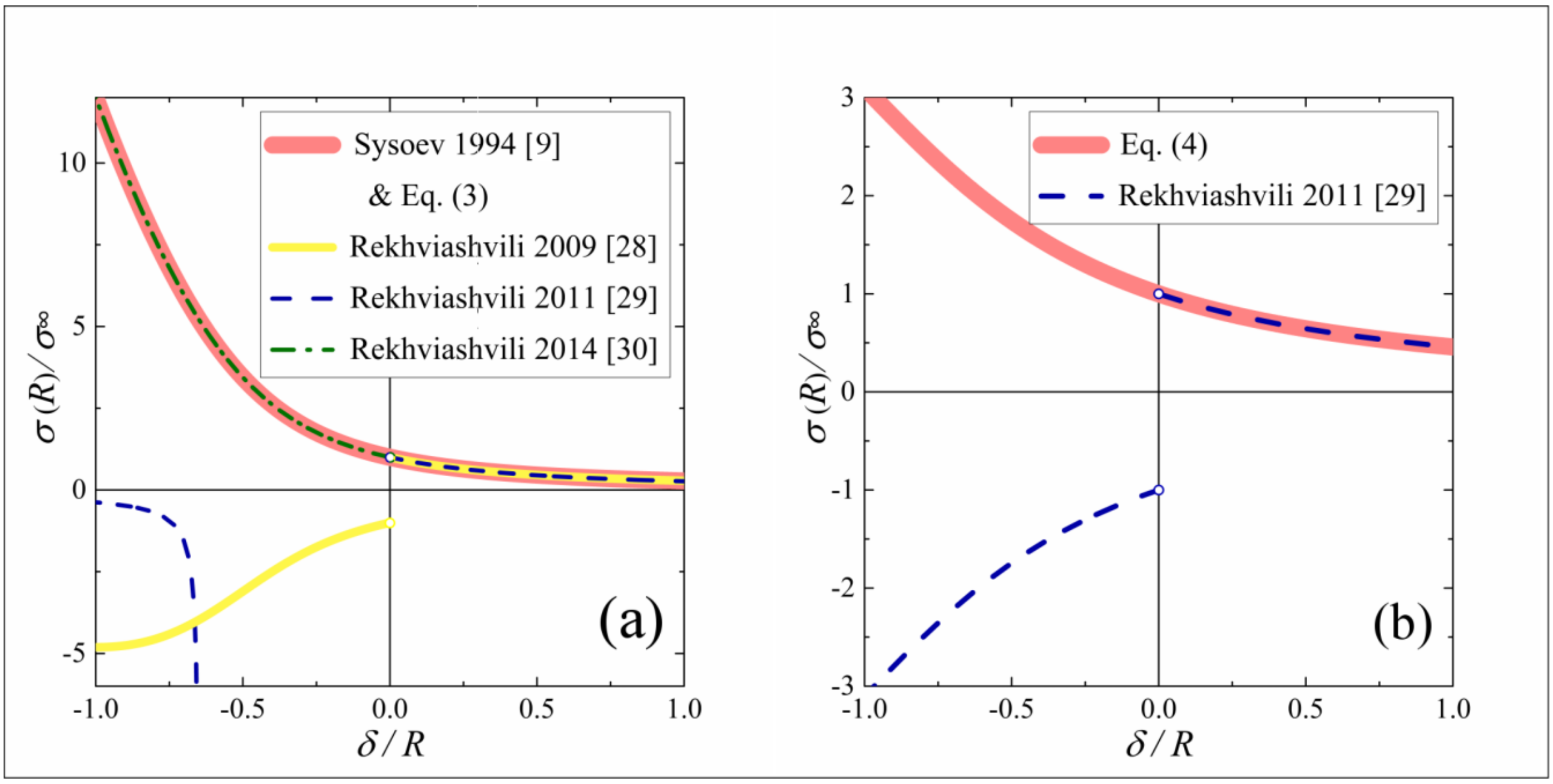

**Fig. 2.** Size dependence of the surface tension of a free surface of an isotropic fluid for the spherical (a) and cylindrical (b) geometries.

The representation of Eqs. (3) and (4) in a Maclaurin series for the small parameter $\delta/R$ gives a well-known approximation for the size dependence of surface tension:

$$\sigma_{s,c}(R)=\sigma^{\infty}\left(1-(n-1)\frac{\delta_{s,c}}{R}+\frac{3}{4}(n-1)^2\left(\frac{\delta_{s,c}}{R}\right)^2-\right.$$
$$\left.-\frac{1}{36}(n-1)^2(17n-19)\left(\frac{\delta_{s,c}}{R}\right)^3+o\left(\frac{\delta_{s,c}}{R}\right)^4\right). \tag{5}$$

## 3. Tolman length evaluation with the use of thermal equation of state

As it was shown in the previous section, for the evaluation of the size dependence of the surface tension, one should know the Tolman length. Since this quantity is on the same order of magnitude as intermolecular distances, its experimental evaluation is extremely difficult [5,23,40,41]. Thus one should find a way which does not require a direct measurement of this quantity for the evaluation of the size dependence of the surface tension.

We use a similar approach based on the thought experiment as proposed earlier in [42,43] for the Tolman length evaluation. We consider an infinitely system, which is equilibrated under temperature $T$ and pressure $P_0$. The pressure was chosen to satisfy a condition of coexistence of the vapor and the liquid phases, and its value was taken from the coexistence curve for a flat surface $P_0 = P_{sat}(T)$. We cut out the spherical and cylindrical volume with the radius $R_0$ from the system and placed it in a medium with saturated vapor with the same temperature. The radius of the volume will be changed to some value $R$ because of a surface tension action. This will lead to the change of the pressure to value $P$, which can be evaluated through the thermal equation of the state ($P = f(V)$). In the frame of thermodynamic approximations, the thermal equation is a first (*zero*) degree homogeneous function of the extensive (*intensive*) parameters. Therefore, one can write the following equation:

$$P - P_0 = f\left(\frac{V_0 - V}{V_0}\right), \tag{6}$$

where $V_0 = V(P_0)$ and $V = V(P)$ are, the initial and resulting volumes, respectively.

For spherical and cylindrical shapes, the argument of the function f can be written as:

$$\frac{V_0 - V}{V_0} = 1 - \left(1 + \frac{\xi}{R}\right)^{-n}, \tag{7}$$

where $\xi = R_0 - R$ ($R >> \xi > 0$), $n = 2$ or 3 for the cylindrical or spherical droplet.

Since $\xi/R$ is an extremely small quantity, the right-hand side of Eq. (6) can be expand in the Maclaurin series

$$P - P_0 = f(0) + nf'(0)\frac{\xi}{R} + \frac{1}{2}\left(-nf'(0) - n^2 f'(0) + n^2 f''(0)\right)\left(\frac{\xi}{R}\right)^2 + o\left(\frac{\xi}{R}\right)^3. \tag{8}$$

The first term in the following equation is equal to zero. On the other hand, one can approximately evaluate surface tension based on the Laplace equation:

$$\Delta P = \frac{n-1}{R}\sigma_{s,c}(R). \tag{9}$$

In the case when the pressure of the saturation vapor is negligible, one can equate the left-hand sides of Eqs. (8) and (9). With the use of the expressions for the surface tension (5) and for the curvature we obtain the following power series:

$$\begin{aligned} & nf'(0)\xi\frac{1}{R}+\frac{1}{2}\left(-nf'(0)-n^2f'(0)+n^2f''(0)\right)\xi^2\frac{1}{R^2}+o\left(\frac{1}{R}\right)^3 = \\ & =(n-1)\sigma^\infty\frac{1}{R}-(n-1)^2\sigma^\infty\delta_{s,c}\frac{1}{R^2}+o\left(\frac{1}{R}\right)^3. \end{aligned} \tag{10}$$

The matching of the multipliers of the terms with $1/R$ and $1/R^2$ on both sides gives the united equation of the Tolman length $\delta$ in the case of the spherical [42] and the cylindrical surfaces:

$$\delta_{s,c} = -\frac{\sigma^\infty}{2}\left(\frac{f''(0)}{\left(f'(0)\right)^2}-\frac{1+n}{n}\frac{1}{f'(0)}\right). \tag{11}$$

The obtained equation can also be written in terms of isothermal compressibility $\beta_T = -\left(\partial V/\partial P\right)/V\big|_{V=V_0,P=P_{sat}}$:

$$\delta_{s,c}(T) = \frac{\sigma^\infty(T)}{2}\left(\frac{1}{\beta_T(T,P)}\left(\frac{\partial\beta_T(T,P)}{\partial P}\right)+\frac{1+n}{n}\beta_T(T,P)\right)\Bigg|_{P_{sat}}. \tag{12}$$

The isothermal compressibility can be represented through the first and second derivatives of the chemical potential $\mu(T,P)$ as a function of pressure:

$$\beta_T(T,P) = -\left(\frac{\partial^2\mu(T,P)}{\partial P^2}\right)\Big/\left(\frac{\partial\mu(T,P)}{\partial P}\right). \tag{13}$$

Consequently, the united equation of the Tolman length $\delta$ for the cylindrical and spherical surfaces of tension can be evaluated as follows:

$$\delta_{s,c}(T) = \frac{\sigma^\infty(T)}{2}\left(\frac{\mu'''(T,P)}{\mu''(T,P)}-\frac{1+2n}{n}\frac{\mu''(T,P)}{\mu'(T,P)}\right)\Bigg|_{P_{sat}}. \tag{14}$$

Thus, our proposed generalized equation of the Tolman length $\delta$ can be applied to both spherical and cylindrical surfaces in a similar manner. As one can see, only the temperature dependence of the surface tension of a flat surface and the first three derivatives of the chemical potential are necessary to evaluate the Tolman length. This information can be taken from the literature for a wide variety of different fluids. Let us emphasize that we have started from the assumption of the homogeneity of the thermal equation. Nevertheless, the size dependence of the surface tension is evaluated based on the perturbation of the initial system's geometry.

## 4. Size dependence of the surface tension based on the data of IAPWS

Let us further consider the system water-vapor in the thermodynamic equilibrium for numerical evaluation of the temperature dependences of the surface tension. The $\sigma^\infty(T)$ and the chemical potential $\mu(T,P)$ of water for a flat surface are taken from IAPWS [44,45]. The derivatives of the chemical potential in a wide range of temperatures and pressures can be obtained from the approximation equation of the specific Gibbs free energy.

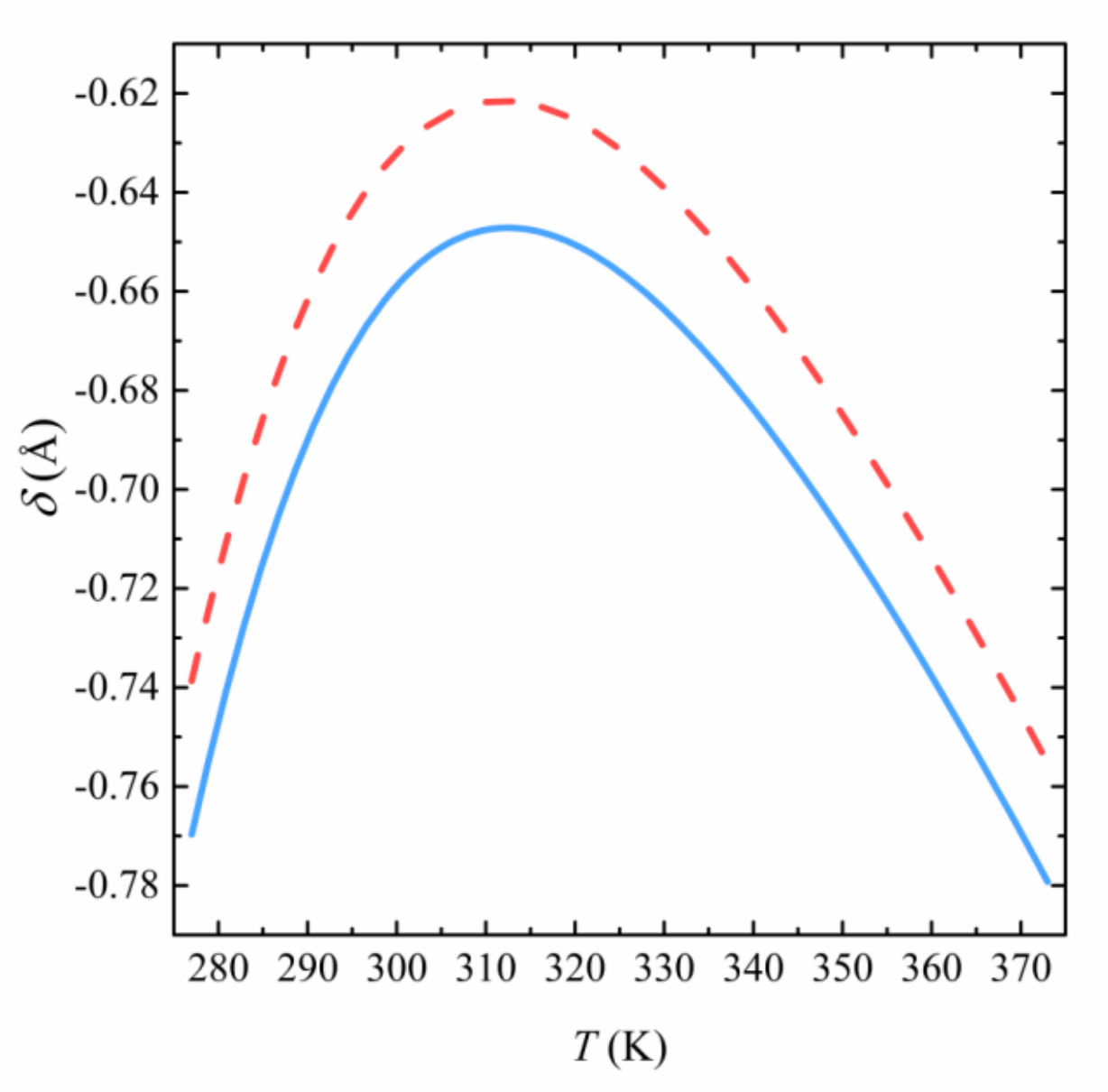

**Fig. 3.** Temperature dependences of the Tolman length $\delta$ for the spherical (solid line) and cylindrical (dashed line) surfaces of tension on the saturation line.

The temperature dependences of the Tolman lengths calculated with Eq. (14) for the cylindrical and spherical droplets in the temperature range from 277 to 373 K are presented in Fig. 3. The dependences are not monotonic and they have well defined maxima equal to 311.83 and 312.45 K for the cylindrical and spherical droplets, respectively. The difference between Tolman lengths for the cylindrical and spherical droplets are approximately 3%-4 % in the considered temperature range. Thus, there is the possibility to evaluate size dependencies of the surface tension as a function of the temperature. As an example, in Fig. 4 the dependencies of the surface tension on the curvature radius for the cylindrical and spherical droplets are depicted for the temperature 291.84 K. This temperature corresponds to the maximum difference of the Tolman length for the two considered cases. For a nanodroplet radius of 1 nm, the surface tension of the cylindrical droplet is approximately 6 % bigger than the surface tension of the spherical one. However, the droplets have approximately the same surface tension if their radii are bigger than 100 nm.

The Tolman length for the cylindrical and spherical droplets were found to be equal to $-0.632 \pm 0.003$ Å and to $-0.659 \pm 0.003$ Å, respectively, at 300 K. The Tolman length for the spherical droplet is in good agreement with experimental values ($\delta = -0.47$ Å [35]) and with molecular dynamics data $\delta = -0.56 \pm 0.09$ Å [18], $\delta = -0.56 \pm 0.1$ Å [20], and $\delta = -0.74$ Å [46].

The resulting dependences of the surface tension $\sigma(R)$ on the radius of curvature of separating surface at 300 K for spherical and cylindrical bubbles (droplets) are presented in Fig. 4 by the dotted and the dashed lines (the dashed-dotted and the dashed-dotted-dotted lines) respectively. The solid line corresponds to the surface of tension of the flat surface. As one can see in Fig. 4, the surface tension of water for the case of the droplet decreases with increasing the droplet size, as opposed to bubbles, for which the surface tension rises with increasing bubble size. Moreover, the rate of change of the surface tension is different for droplets and bubbles.

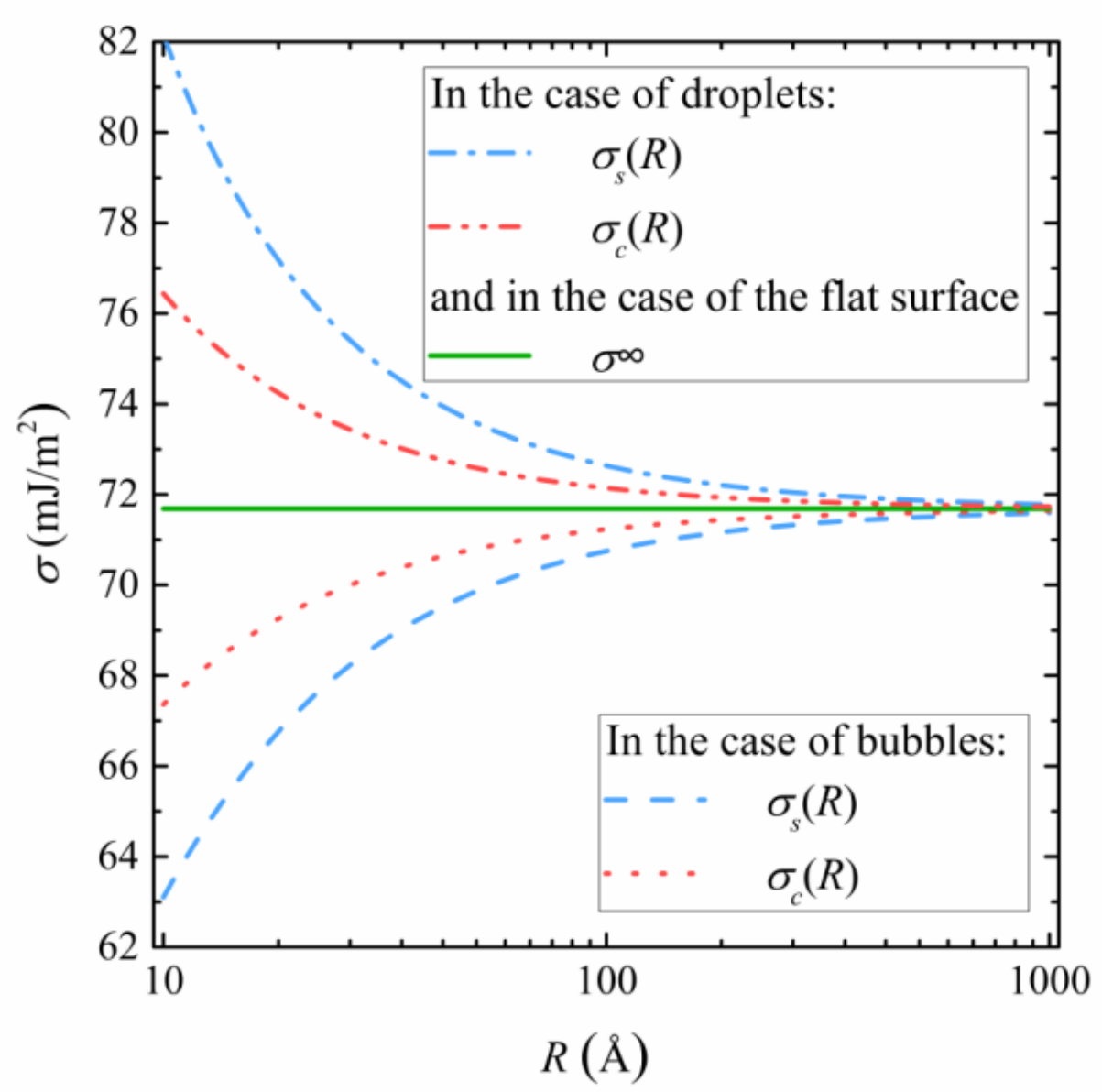

**Fig. 4**. Dependences of the surface tension $\sigma(R)$ on the radius of curvature of separating surface at 300 K for spherical and cylindrical bubbles (droplets) are presented by the dotted and the dashed lines (the dashed-dotted and the dashed-dotted-dotted lines) respectively. The solid line corresponds to the surface of tension of the flat surface.

Here we note that the obtained expressions for the Tolman length (11), (12), and (14) are valid only in the approximation, when $\xi = R_0 - R << R$. The validity of this condition in the examined temperature range can be estimated on the saturation line in the case of spherical ($n = 3$) and cylindrical ($n = 2$) "droplets" of water:

$$\frac{1-n}{n}\sigma^{\infty}(T,P)\frac{\mu''(T,P)}{\mu'(T,P)}\bigg|_{P_{sat}} << R. \tag{15}$$

Therefore, this approximation is correct in the examined temperature range on the saturation line for nanoscopic droplets of the range $10^{-9}$ – $10^{-7}$ m.

## 5. Conclusions

In conclusion, we considered and evaluated the GTKB equation on the assumption that Tolman length is independent of the curvature. An analytical solution of the GTKB equation for the cases of cylindrical and spherical surfaces of the tension without any assumptions regarding sign of $\delta/R$ parameter was proposed. Our approach allows evaluating the size dependence of the surface tension in the same manner for $\delta > 0$ and $\delta < 0$ ($R > 0$ and $R < 0$) alike with the use of the data of the International Association for the Properties of Water and Steam. It is shown that for both cases of the Tolman length signs, the solution is a continuous function for all positive and negative values, including the case of zero curvature. The use of the Maclaurin representation for $\delta/R$ of these solutions allows reformulating the equations for the Tolman length estimation. The Tolman length for the "water saturated vapor" interface with our methodology is in good agreement with molecular dynamics and experimental data.